\input{aipcheck}

\documentclass[
    ,final            
  ]
  {aipproc}

\layoutstyle{6x9}

\def \civ {C\,{\textsc {iv}}}

\def \oii {[O\,{\textsc {ii}}]}
\def \oiii {[O\,{\textsc {iii}}]}

\let\oldsqrt\sqrt
\def\sqrt{\mathpalette\DHLhksqrt}
\def\DHLhksqrt#1#2{
\setbox0=\hbox{$#1\oldsqrt{#2\,}$}\dimen0=\ht0
\advance\dimen0-0.3\ht0
\setbox2=\hbox{\vrule height\ht0 depth -\dimen0}
{\box0\lower0.4pt\box2}}

\begin{document}

\title{Finding SDSS BALQSOs Using Non-Negative Matrix Factorisation}


\author{James T. Allen}{
  address={Institute of Astronomy, University of Cambridge, Madingley
  Road, Cambridge CB3 0HA}
}

\author{Paul C. Hewett}{
  address={Institute of Astronomy, University of Cambridge, Madingley
  Road, Cambridge CB3 0HA}
}

\author{Vasily Belokurov}{
  address={Institute of Astronomy, University of Cambridge, Madingley
  Road, Cambridge CB3 0HA}
}

\author{Vivienne Wild}{
  address={Max-Planck-Institut f\"ur Astrophysik, Karl-Schwarzschild
  Str. 1, 85741 Garching, Germany}
}

\begin{abstract}
Modern spectroscopic databases provide a wealth of information about
the physical processes and environments associated with astrophysical
populations.  Techniques such as blind source separation (BSS), in
which sets of spectra are decomposed into a number of components,
offer the prospect of identifying the signatures of the underlying
physical emission processes.  Principle Component Analysis (PCA) has
been applied with some success but is severely limited by the inherent
orthogonality restriction that the components must satisfy.

Non-negative matrix factorisation (NMF) is a relatively new BSS
technique that incorporates a non-negativity constraint on its
components.  In this respect, the resulting components may more
closely reflect the physical emission signatures than is the case
using PCA.  
We discuss some of the considerations that must be made when applying
NMF and, through its application to the quasar spectra in the Sloan
Digital Sky Survey (SDSS) DR6, we show that NMF is
a fast method for generating compact and accurate reconstructions of
the spectra.

The ability to reconstruct spectra accurately has numerous
astrophysical applications.  Combined with improved SDSS redshifts, we
apply NMF to the problem of defining robust continua for quasars that
exhibit strong broad absorption line (BAL) systems.   The resulting
catalogue of SDSS DR6 BAL quasars will be the largest available.
Importantly, the NMF approach allows quantitative error estimates to be 
derived for the Balnicity Indices as a function of key astrophysical and
observational parameters, such as the quasar redshifts and the 
signal-to-noise ratio of the spectra.

%
%

\end{abstract}

\maketitle


\section{Introduction}

The scale of modern spectroscopic surveys such as the Sloan Digital
Sky Survey (SDSS) demand the development and application of new analysis
techniques that are able to condense the vast quantities of available
data to extract useful results.  Blind source separation (BSS)
techniques are a family of techniques that enable such a condensation
of information.  A BSS technique can take a set of spectra written as
a matrix, $V$, and generate a corresponding set of components, $H$,
that can be linearly combined to recreate the original data using a
set of weighting coefficients, $W$:
\begin{equation}
V = WH.
\end{equation}
The form of the components depends on the particular technique used.

The most widely used BSS technique in astronomy is principal component
analysis (PCA), which generates orthogonal components.  PCA has been
successfully applied for a number of purposes but interpretation of
the component spectra is limited by their orthogonality.  In this work
we describe non-negative matrix factorisation, a recently-developed
BSS technique, and demonstrate its application to the reconstruction
of absorbed quasar continua in the SDSS.

\section{Non-Negative Matrix Factorisation}

\subsection{Definition}

Non-negative matrix factorisation (NMF) is a relatively new BSS
technique that incorporates a non-negativity constraint on both its
components and their weights \cite{LS99,LS00,BR07}.  The
non-negativity constraint is appealing in the context of spectroscopic
data as the physical emission signatures are expected to naturally
obey this restriction.  Unusually for a BSS technique, fewer
components are generated than there are input spectra.  Starting from
random initial matrices, the components ($H$) and weights ($W$) follow
the multiplicative update rules
\begin{equation}
\label{eq:wupdate}
W_{ik} \leftarrow W_{ik} \frac{[VH^T]_{ik}}{[WHH^T]_{ik}},
\end{equation}
and
\begin{equation}
\label{eq:hupdate}
H_{kj} \leftarrow H_{kj} \frac{[W^TV]_{kj}}{[W^TWH]_{kj}}.
\end{equation}

As fewer components are generated than there are
input spectra the reconstructions $WH$ will in general be
approximations to the data; the update rules minimise the error in the
approximations, as measured by the Euclidean distance between the
reconstructions and the data.  The random starting conditions result
in slightly different components being generated each time the
algorithm is executed, but the resulting reconstructions do not vary.

When components and weights have been generated from one set of input
spectra, the components can be applied to generate reconstructions of
other spectra.  In this case random initial weights are used, which
are updated according to equation~\ref{eq:wupdate}, while the
components are held fixed.

%
%
\subsection{Practicalities of Applying NMF}

\subsubsection{Number of Components}

In applying NMF to any dataset the number of components must be
pre--specified.  Increasing the number of components will always
increase the precision of the reconstructions, but the increased
precision is not always desirable.  With too many components it 
becomes beneficial for the NMF procedure, in terms of
the total error in the reconstructions, to overfit a
small number of particularly noisy or very unusual spectra, incorporating
their noise, or their unique features, in the components. It is thus
most effective to employ the maximum number of components that does not 
produce overfitting in any subset of spectra. The optimal number of
components must be chosen separately for each sample. For reconstructions
of the SDSS quasar spectra the number is between 8 and 15, with the 
appropriate number determined via a simple trial and error scheme.

\subsubsection{Sample Selection}

Some care must be taken when selecting the sample to use as inputs to
the NMF algorithm.  A larger sample will better constrain the
resulting components, but this must be balanced against the increased
CPU time required for the calculations.  A sample size of around 500
has been found to produce well-constrained components while completing
in a reasonable time on a modern desktop computer.

The range of properties of the sample spectra should reflect the range
of properties of the population from which they are drawn.  Individual
objects with very unusual properties can cause problems by inducing
overfitting at a smaller number of components than would otherwise be
the case.  In such cases the most extreme objects must be removed from
the sample.

\subsubsection{Redshifts}

Before a source separation can be performed the input spectra must be
shifted to their respective rest frames.  Accurate
redshifts are required to ensure that specific
features occur in the same location in all spectra.  Redshift errors
of $\sim0.001$ are sufficient to blur
narrow features when viewed across the sample, reducing the quality of
the NMF reconstructions.

In the following work the redshifts used were recalculated from the
SDSS spectra in order to reduce the errors and biases present in the SDSS
redshifts, which are each as large as $\pm0.005$ at redshifts $z > 2$.
At low redshifts the new values were calculated from the positions
of the \oiii\ and \oii\ emission lines, while at higher redshifts
improved cross-correlation measures were developed that are unbiased
with respect to the narrow emission line redshifts at low-redshift.

\subsubsection{Dust Reddening}

NMF reconstructs the observed spectra as a linear combination of components,
assuming all contributions are additive.  In
contrast, dust reddening multiplies the spectra by a
wavelength-dependent flux ratio.  For moderate levels of reddening
this effect is smaller than the natural object-to-object variations
between spectra and is accounted for in the NMF components, but
this is not the case for the most heavily obscured objects.

To improve the quality of the reconstructions of heavily reddened
objects an estimate of the unreddened spectrum can be made.  By
multiplying the observed spectrum by a power law slope it can be
given a slope within the typical range for the object being observed.
The edited spectrum is then used as an input for the NMF algorithm,
and the results can be divided by the same power law slope to match
the observed spectrum.

\section{Application to Broad Absorption Line Quasars}

\subsection{Broad Absorption Line Quasars}

Broad absorption line quasars (BALQSOs) are quasars that exhibit
strong broad absorption line (BAL) systems, with velocity widths often
in excess of 10~000~km~s$^{-1}$.  They are always intrinsic to the
quasar, although they may be blueshifted with respect to the active
galactic nucleus (AGN) by
over 20~000~km~s$^{-1}$.  The blueshift is taken to mean that the BAL
systems are part of large-scale outflows driven by the AGN.

Methods of classifying BALQSOs vary, but the most widely used measure
is the balnicity index (BI) \cite{WMFH91}, defined as
\begin{equation}
\label{eq:bi}
{\rm BI} = - \int_{25000 {\rm \ km\ s}^{-1}}^{3000 {\rm \ km\ s}^{-1}}
\left( 1 - \frac{f(v)}{0.9} \right) C {\rm d}v,
\end{equation}
where $f(v)$ is the continuum-normalised flux as a function of
velocity, $v$,  relative to the line centre.  The constant $C$ is equal to 1
in regions where $f(v)$ has been continuously less than 0.9 for at
least 2000~km~s$^{-1}$, and 0 elsewhere.

Depending on the measure
used, the observed BAL fraction is between 10 and 15
per cent \cite{WMFH91,Trump06,Knigge08}; the fraction increases to 17
to 22 per cent when corrected for differential selection effects
between BAL and non-BAL quasars.  The presence of BAL systems in some,
but not all, quasars could be the result of an orientation
effect \cite{WMFH91,SH99}, or BALQSOs could represent a particular
stage in the quasar life-cycle \cite{Voit93,Becker00}.

\subsection{Application of NMF}

In order to characterise BALQSOs, estimates of the unabsorbed
continua are required, which we have generated using the NMF
technique.  Samples of spectra, each consisting of 500 non-BAL quasars, was
selected to cover all redshifts, $0.7 < z < 2.6$, in redshift bins of
width $\Delta z = 0.1$. The NMF algorithm was applied to each sample
to produce NMF component spectra, applicable to quasars within a specified
redshift interval.

The resulting components were used to reconstruct the continua of
potential BALQSO spectra.  During this fitting the components were
held fixed and only the weights were updated, following
equation~\ref{eq:wupdate}.  All regions where broad absorption is
likely to occur were initially masked, and the mask was then
iteratively updated by comparing the resulting reconstructed continuum
with the observed spectrum.  The NMF fitting was recalculated after
each mask update; in most cases the mask locations reached a stable
solution after only two or three iterations.

The classification of quasars into BAL and non-BAL categories is
sensitive to the redshift used in the calculations, as an inaccurate
redshift can move absorption features into or out of the velocity
range examined.  An inaccurate redshift will also reduce the
quality of any estimate of the continuum.  Unfortunately, inaccurate
redshifts are more likely for BALQSOs, as the absorption of the blue
wing of the \civ\ line means that any method using this line will
overestimate the redshift.  The redshift measurements used here do not
normally use the regions of the spectra that exhibit broad absorption,
so they are not expected to be significantly affected by this bias.

At redshifts $z \ge 2.6$ there are insufficient non-BAL quasars with
adequate signal-to-noise ratio (SNR) available to produce useful components.  
However, due to the Ly$\alpha$ forest, no additional quasar ``continuum''
enters the spectra at these redshifts so the components from 
the $2.5 \le z < 2.6$ redshift bin were used to fit quasars 
with $z \ge 2.6$.

\subsection{Results}

\begin{figure}
  \includegraphics[height=.945\textheight]{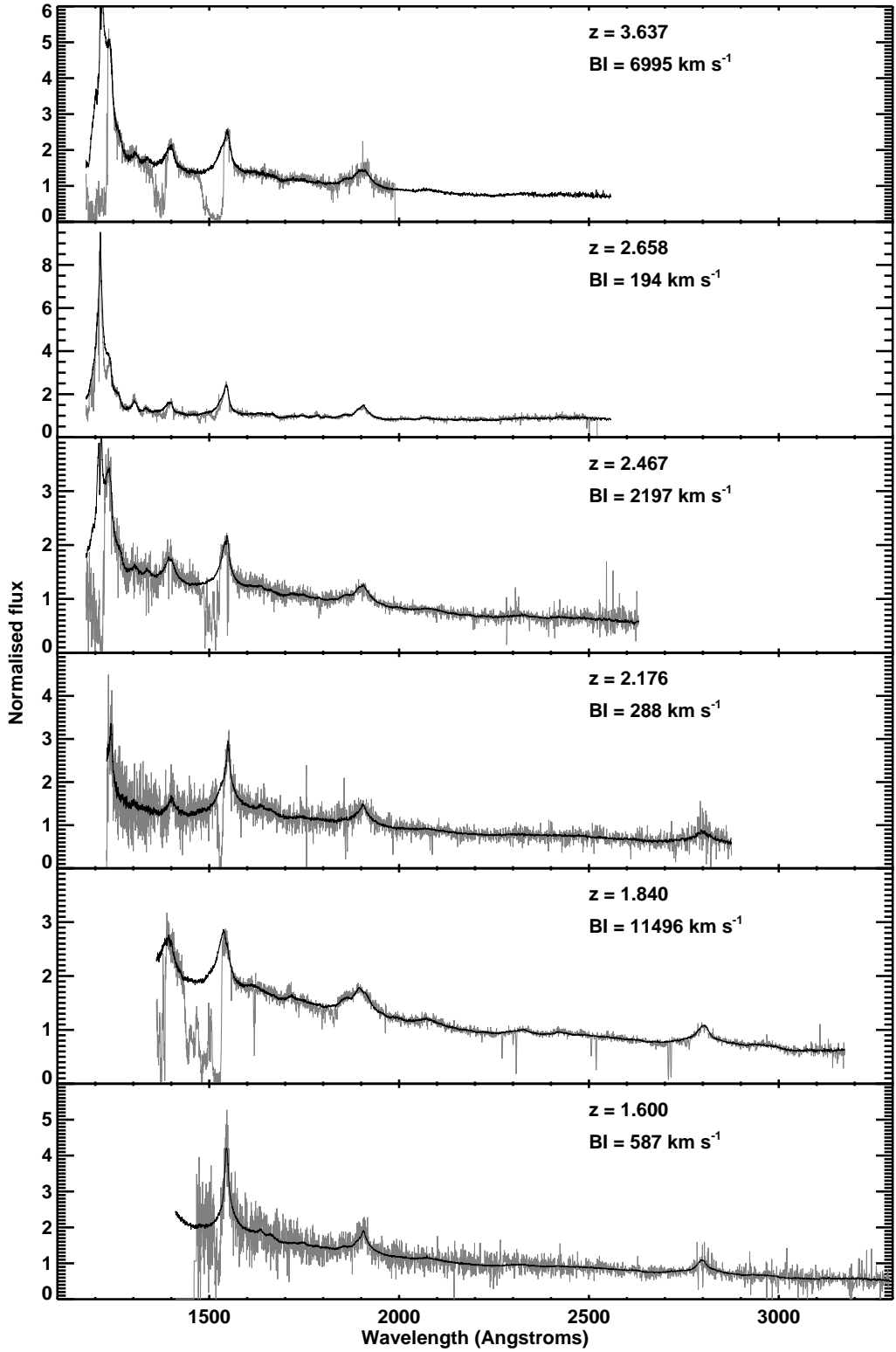}
  \caption{Example reconstructions (black curves) of BALQSO continua,
  based on their observed spectra (grey curves).  Redshifts and
  balnicity indices are as shown in the panels.}
  \label{fig:receg}
\end{figure}

The NMF-procedure described above has been applied to generate estimated
continua for over 80~000 quasar spectra with redshifts $0.7 < z < 4.4$. 
Example reconstructions of BALQSO continua with a range of redshifts
and BI values are shown in
Figure~\ref{fig:receg}.

In order to quantify the accuracy with which the continua are
reconstructed, a set of synthetic BALQSO spectra were created.  
First, continua were fitted to a set of 105 BALQSOs,
redshifts $2.25 < z < 2.35$, with high SNR spectra,
using the procedure outlined above.
A flux-ratio spectrum for each BALQSO was generated by dividing the
observed spectrum by the reconstruction, then smoothing to remove
noise.  A set of randomly chosen, non-BAL quasar spectra were then 
multiplied by the flux ratio spectra, generating synthetic BALQSO spectra 
for which the unabsorbed continua were known.

\begin{figure}
  \includegraphics[height=.3\textheight]{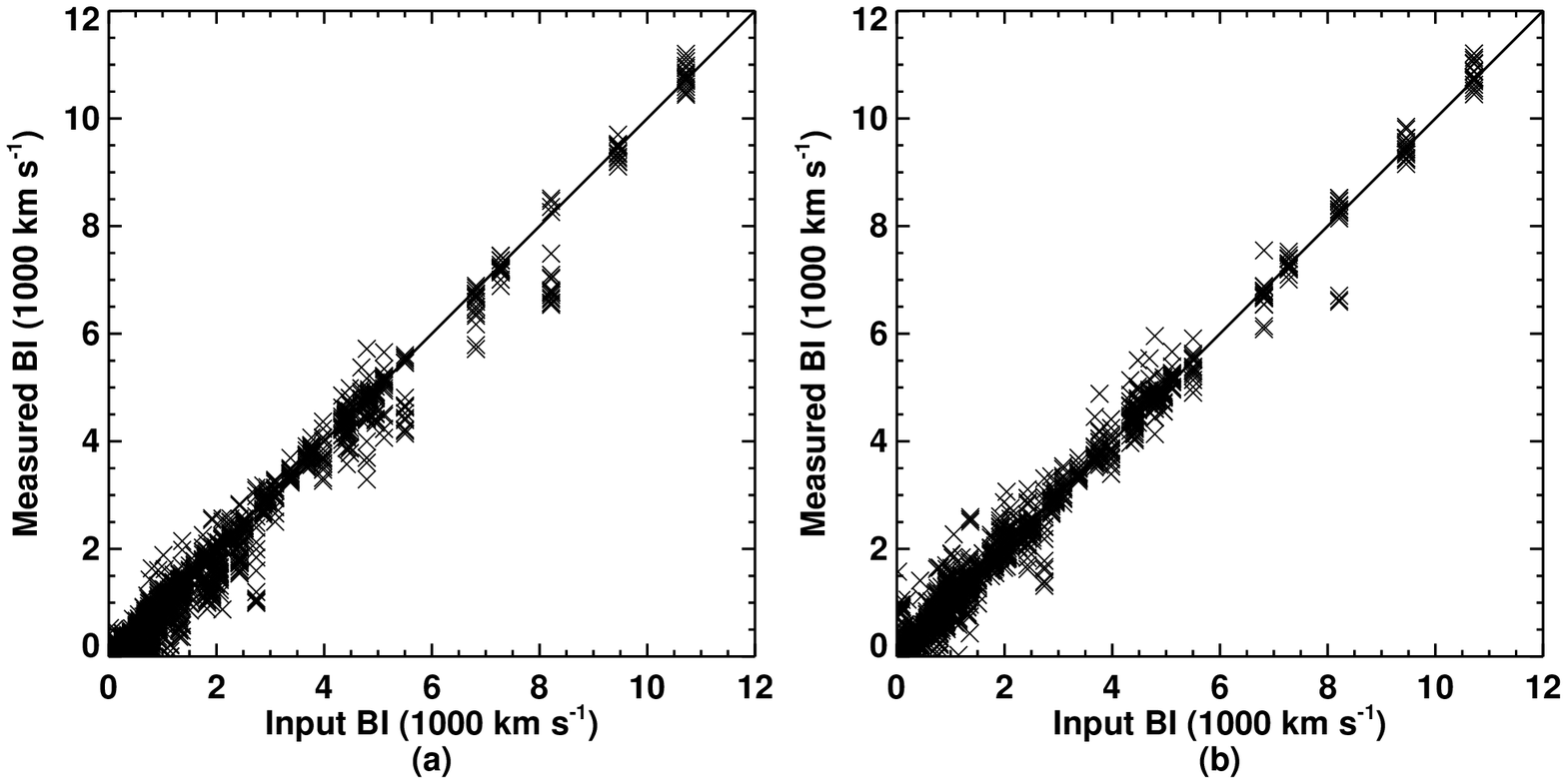}
  \caption{(a) Measured BI values for a set of synthetic BALQSO spectra
  with known input BI.  (b) As (a) but with the flux ratios smoothed
  before the BI is measured.}
  \label{fig:testresults}
\end{figure}

Continua were fitted to the synthetic BALQSO spectra by the same
method as for genuine BALQSOs, and the resulting BI measurements were
compared to the real BI values of the input flux ratios as a
quantitative test of the NMF procedure.  The results are shown in
Figure~\ref{fig:testresults}~(a).  In most cases there is good
agreement between the input and measured values, but the BI is
frequently undermeasured.  The undermeasurement is due largely to 
the effect of noise in the lower SNR spectra: a
single pixel with positive noise that brings its value above 90\% of
the continuum can prevent a large velocity interval from contributing 
to the BI measurement.

Figure~\ref{fig:testresults}~(b) shows the BI values measured when the
noise is reduced by smoothing the measured flux ratios in the same
way as for the input flux ratios.  The smoothing reduces significantly 
the number of systems in which the BI is undermeasured, resulting in an RMS
residual of 250~km~s$^{-1}$ with a mean offset of only 20~km~s$^{-1}$.

The effect of noise in BI calculations has been noted
previously \cite{Knigge08} but to date it has not been quantified.
A forthcoming analysis of the undermeasurement of the BI at different
signal-to-noise ratios will allow an accurate determination of the BAL 
fraction as a function of quasar luminosity and redshift.

\section{Conclusions}

Non-negative matrix factorisation shows much promise as a technique
for analysis of modern spectroscopic surveys such as the SDSS.  We
have demonstrated here its application to the problem of estimating
the continua of BALQSOs.  Component spectra were generated from sets of
non-BAL quasars, and then used in conjunction with iteratively
defined masks to identify regions of absorption in other quasar
spectra and reconstruct their continua.

The effectiveness of the NMF continuum definition has been demonstrated
using simulated BALQSO spectra. The BI values of BALQSOs are often 
undermeasured due to the influence of limited signal-to-noise ratio in
the observed spectra. A forthcoming analysis of this undermeasurement effect
will allow the BAL fraction in the SDSS spectroscopic survey to be 
quantified as a function of key parameters such as quasar redshift and
luminosity.


\begin{theacknowledgments}
James Allen and Vasily Belokurov acknowledge the award of an STFC Ph.D. 
studentship and an STFC Postdoctoral Fellowship respectively. Paul Hewett
acknowledges support from the STFC-funded Galaxy Formation and Evolution
programme at the Institute of Astronomy.
\end{theacknowledgments}



\bibliographystyle{aipproc}   

\bibliography{cpgsbib}

\IfFileExists{\jobname.bbl}{}
 {\typeout{}
  \typeout{******************************************}
  \typeout{** Please run "bibtex \jobname" to optain}
  \typeout{** the bibliography and then re-run LaTeX}
  \typeout{** twice to fix the references!}
  \typeout{******************************************}
  \typeout{}
 }

\end{document}